\documentclass[english]{article}
\usepackage[T1]{fontenc}
\usepackage[latin9]{inputenc}
\usepackage{geometry}
\usepackage{color}
\usepackage{amssymb}
\usepackage{stackrel}
\usepackage{graphicx}
\PassOptionsToPackage{normalem}{ulem}
\usepackage{ulem}

\makeatletter
\usepackage{times,cite}
\date{}

\makeatother

\usepackage{babel}
\begin{document}
\title{Are optical quantum information processing experiments possible without
beamsplitter?}
\author{{\normalsize{}Kishore Thapliyal$^{\ddagger,}$}\thanks{Email: kishore.thapliyal@upol.cz }{\normalsize{}
~ and Anirban Pathak$^{\mathsection,}$}\thanks{Email: anirban.pathak@jiit.ac.in}{\normalsize{}
}\\
{\normalsize{}$^{\ddagger}$RCPTM, Joint Laboratory of Optics of Palacky
University and Institute of Physics}\\
{\normalsize{}of Academy of Science of the Czech Republic, Faculty
of Science, Palacky University, }\\
{\normalsize{} 17. listopadu 12, 771 46 Olomouc, Czech Republic}\\
{\normalsize{}$^{\mathsection}$Jaypee Institute of Information Technology,
A-10, Sector-62, Noida, UP-201309, India}}
\maketitle
\begin{abstract}
The significance of beamsplitter in experimental optical quantum information
processing and quantum technology is discussed with
a focus on the role of a beamsplitter-type Hamiltonian in the recent
development in this field of research. Here, we follow
a new approach to briefly describe quantum measurement, Bell measurement,
quantum state engineering, quantum teleportation, cryptography, and
computation using both discrete and continuous variables
to establish the wide applications of beamsplitter-type operation.
Finally, we also discuss the limitations of this linear optical element.

\end{abstract}

\section{Introduction}

As reflected from the title page of this issue of the journal, all
the articles of this issue are dedicated to Prof.\ Ajoy Ghatak who
has just become an octogenarian. Many of us (including the authors of
this work) have learned the basic ideas of optics and quantum mechanics
from the excellent books \cite{ghatak1998introduction,ghatak2004quantum,ghatak2005optics,ghatak2015experiments}
authored by him. A characteristic of his books that mesmerized us
over the years was their simplicity. Motivated by that and the fact
that most of his research works \cite{diggavi1989perturbation,thyagarajan2011quantum}
and books \cite{ghatak1998introduction,ghatak2004quantum,ghatak2005optics,ghatak2015experiments}
involve traditional optics, fiber optics, and quantum mechanics, we
planned to write this article on the modern applications of a very
simple component that connects all the three domains of his interest.
Specifically, we want to focus this paper on beamsplitter (BS) and its modern applications.
BS is a well-known and simple linear optical component that every
interferometer contains, be it a simple Michelson interferometer
(MI) described in Chapter 15 of Prof.\ Ghatak's famous book entitled
Optics \cite{ghatak2005optics}, or a more sophisticated version of
MI used in the famous LIGO experiment to detect gravitational wave \cite{aasi2013enhanced,grote2013first};
be it a single photon-based Mach-Zehnder interferometer (MZI) that
can be used to establish the existence of quantum superposition and
collapse on measurement postulate of quantum mechanics \cite{pathak2019light} or a nested version of MZI used in the recent proposals for counterfactual
quantum communication \cite{salih2013protocol}. In fact, any piece
of glass can be viewed as a BS. Of course, it will not be a 50:50
(symmetric) BS, but will indeed be a BS. For example, a piece of pure
glass which reflects only 4\% of the incident light can be viewed
as 4:96=1:24 (asymmetric) BS.  Further, in what follows, we will see
that optical couplers (which are primary component of the majority of
the optical fiber based experiments and any integrated-optic device)
are equivalent to BS. The relevance of BSs in designing interferometers,
like MI or MZI was known since long, but recently it is realized that
the sensitivity of an MI can be enhanced considerably if squeezed
vacuum is inserted from the free mode of the BS in an MI, and that
enhanced sensitivity can be used to detect gravitational wave \cite{aasi2013enhanced,grote2013first}.
A true random number generator can be built by using a single photon
source (or an approximate version of that which uses weak coherent
pulse) and a BS \cite{herrero2017quantum}. Beyond these applications,
BSs have been observed to be used in most of the fascinating experiments
of quantum optics (e.g., Hanbury Brown-Twiss (HBT) experiment \cite{brown1956correlation,fox2006quantum},
homodyne detection \cite{breitenbach1997measurement,bachor2004guide},
characterization of squeezed \cite{breitenbach1997measurement} and
antibunched \cite{kimble1977photon} states, Bell's inequality \cite{rarity1990experimental,pan2000experimental},
higher-order nonclassicality \cite{avenhaus2010accessing}), quantum
information (quantum teleportation \cite{bouwmeester1997experimental},
densecoding \cite{mattle1996dense}, photon subtraction in decoy state
quantum key distribution \cite{lo2014secure}, measurement device
independent quantum cryptography \cite{ma2018continuous}, continuous
variable quantum cryptography \cite{jouguet2013experimental,pirandola2015high},
cryptanalysis in quantum cryptography \cite{pirandola2008continuous,liu2011proof}),
quantum state engineering (photon subtraction \cite{zavatta2008subtracting,gerrits2010generation,malpani2019lower},
quantum scissors \cite{leonski2011quantum}, and entanglement generation
\cite{zukowski1997realizable}), which will be further discussed in
the following sections. This observation led to the question ``Is
it possible to design an optical quantum information processing experiment
without using a BS?''. We aim to address this question in the remaining
part of this paper. 

The rest of the paper is organized as follows. In Section \ref{sec:Mathematical-modeling},
we introduce the mathematical details of BS operation and its role
in quantum optics and information. Significance of BS in discrete
and continuous variable quantum communication is discussed in Sections
\ref{sec:discrete} and \ref{sec:continuous}, respectively. Subsequently,
the role of BS in discrete and continuous variable quantum computation
is summarized in Section \ref{sec:computation}. Finally, applications
of BS operation in other areas of research in the field of quantum
foundations, quantum information processing, and quantum
technology are discussed in Section \ref{sec:Other} before concluding
the paper in Section \ref{sec:Conclusion}.

\section{Mathematical modeling of beamsplitter and relevance in quantum optics
and information \label{sec:Mathematical-modeling}}

\begin{figure}
\begin{centering}
\includegraphics{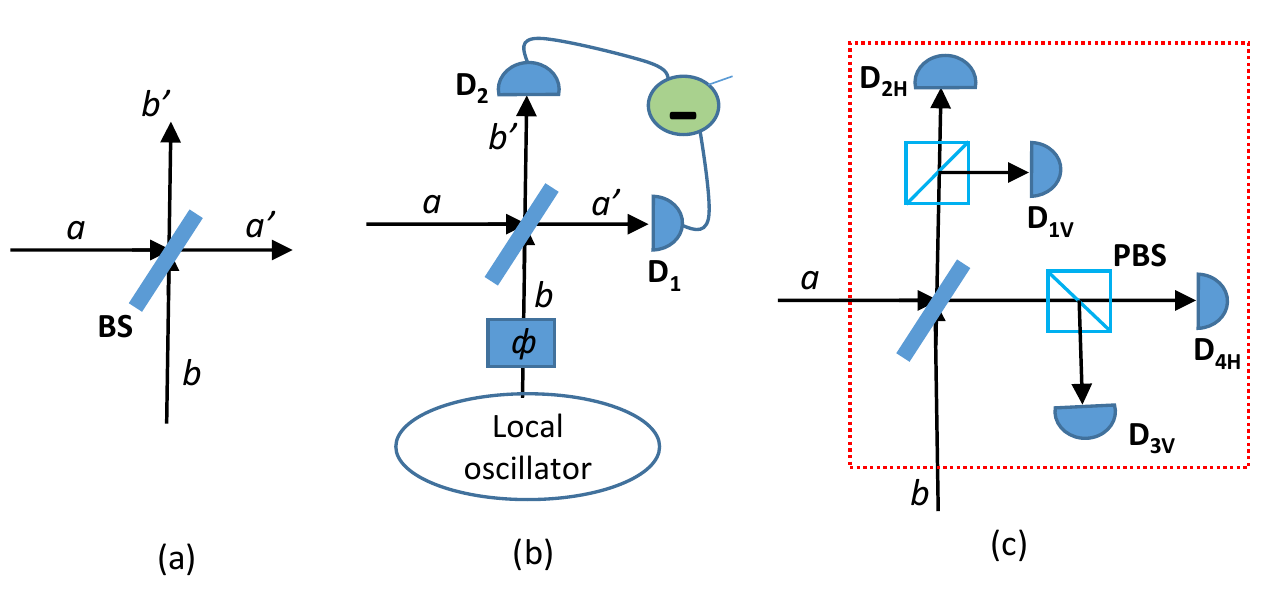}
\par\end{centering}
\caption{\label{fig:BS}(Color online) (a) BS with input-output
relation, (b) homodyne detection with photon detectors ${\rm D}_{i}$,
and (c) Bell measurement (BM) using a BS and two polarizing beamsplitters
(PBSs).}
\end{figure}

A BS is a semitransparent thin film which transmits (reflects) a part
of the incident beam of light of amplitude $E$ with transmission
(reflection) amplitude $t$ $\left(r\right)$, i.e., $tE$ $\left(rE\right)$.
In case of quantized fields, field amplitudes can be denoted by corresponding
field operator $a$. Two output modes of the BS in terms of the input
modes (as shown in Fig. \ref{fig:BS} (a)), reflection and transmission
coefficients can be defined as \cite{agarwal2013quantum}

\begin{equation}
\left(\begin{array}{c}
a^{\prime}\\
b^{\prime}
\end{array}\right)=\left(\begin{array}{cc}
t & r\\
r & t
\end{array}\right)\left(\begin{array}{c}
a\\
b
\end{array}\right)=U_{{\rm BS}}\left(\begin{array}{c}
a\\
b
\end{array}\right),\label{eq:BS}
\end{equation}
where without loss of generality we have assumed transmission and
reflection amplitudes are the same for both the inputs of the BS.
Here, $a$ $\left(a^{\dagger}\right)$ and $b$ $\left(b^{\dagger}\right)$
correspond to the annihilation (creation) operators of two modes of
the BS. The requirement for the validity of commutation relation $\left[A_{i},A_{j}^{\dagger}\right]=\delta_{ij}$
is same as the conservation of energy on a lossless BS, $\left|t\right|^{2}+\left|r\right|^{2}=1$
and $t^{*}r+r^{*}t=0$. Using these conditions, we can parameterize
transmission and reflection coefficients as $t=\cos\theta$ and $r=\sin\theta\exp\left(i\Phi\right)$.
In the present work, we assume $\Phi=\frac{\pi}{2}$. Thus, $a^{\prime}=a\left(\theta\right)$
and $b^{\prime}=b\left(\theta\right)$ in Eq. (\ref{eq:BS}) can be
interpreted as the solution of differential equations that can be
incidentally interpreted as Heisenberg's equations of motion with
the effective Hamiltonian 
\begin{equation}
\begin{array}{lcl}
H_{{\rm BS}} & =-\hbar & \left(a^{\dagger}b+ab^{\dagger}\right).\end{array}\label{eq:ham}
\end{equation}
Thus, a unitary operator which represents a BS can be defined as
\begin{equation}
\begin{array}{lcl}
U_{{\rm BS}} & = & \exp\left\{ i\theta\left(a^{\dagger}b+ab^{\dagger}\right)\right\} .\end{array}\label{eq:Ubs}
\end{equation}

Note that the same Hamiltonian (\ref{eq:ham}) also describes another
optical system, namely linear optical directional coupler. In fact,
it describes a family of physical systems of practical relevance.
For example, it describes an atom-atom two-component BEC system \cite{mewes1997output,torii2000mach}.
However, here we wish to restrict to optical systems and note that
the linear optical coupler forms an integral part of the integrated
waveguide system used in optical quantum information processing experiments.

\subsection{Role of beamsplitter in quantum optics and measurments }

Homodyne measurements allow us to measure one of the quadrature variables
$X=\frac{1}{2}\left(a+a^{\dagger}\right)$ and $Y=\frac{1}{2i}\left(a-a^{\dagger}\right)$
analogous to dimensionless position and momentum in the classical
phase space. It involves mixing the single-mode to be measured with
a strong classical coherent field $\left|\left|\alpha\right|\exp\left(i\phi\right)\right\rangle $
(called local oscillator) at a BS as input modes $a$ and $b$, respectively
(cf. Fig. \ref{fig:BS} (b)). The difference in the output currents
of the BS can be defined in terms of output $a^{\prime}$ and $b^{\prime}$
as $\langle a^{\prime\dagger}a^{\prime}-b^{\prime\dagger}b^{\prime}\rangle=i\langle a^{\dagger}b-b^{\dagger}a\rangle.$
This can be simplified for coherent field initially in mode $b$ as
$2|\alpha|\left\langle X\cos\left(\phi+\frac{\pi}{2}\right)+Y\sin\left(\phi+\frac{\pi}{2}\right)\right\rangle .$
Notice that by choosing $\phi=-\frac{\pi}{2}$ and $\phi=0$ we can
measure quadrature $X$ and $Y$, respectively. On top of that, by
choosing different values of parameter $\phi$ marginal distributions
along rotated quadrature in the phase space can be measured. Repeated
measurement of such marginals with corresponding $\phi$, known as
optical tomography \cite{thapliyal2016tomograms}, allows one to reconstruct
a distribution function in the phase space, i.e., Wigner function
\cite{breitenbach1997measurement}. In fact, BS plays a significant
role in reconstructing the Wigner function, characterization of entanglement
and steering as well \cite{bachor2004guide}. Further, homodyne measurement
is important in the continuous variable quantum information processing. 

Discrete variable quantum communication and computation (specially
BB84 types schemes of quantum key distribution and other cryptographic
tasks \cite{bennett1984quantum,gisin2002quantum,sharma2016comparative})
often desire a source which can generate single photon at will. Characterization
of such sources of light are based on HBT experiment, which measures
second-order intensity correlation defined as $g^{(2)}(\tau)=\frac{\langle n_{a^{\prime}}(t)n_{b^{\prime}}(t+\tau)\rangle}{\langle n_{a^{\prime}}(t)\rangle\langle n_{b^{\prime}}(t+\tau)\rangle}$,
where $n_{A}=A^{\dagger}A$ is the number operator. It corresponds
to the detection of a photon in the output mode $a^{\prime}$ of the
beamplitter at time $t$ followed by a photon detected in mode $b^{\prime}$
at time $t+\tau$, which is normalized such that $g^{(2)}(\tau)=1$
for a coherent state. The light is antibunched if $g^{(2)}(0)<g^{(2)}(\tau)$.
A detailed discussion of antibunching can be found in our recent works
\cite{ghatak2017light,thapliyal2014higher,thapliyal2017comparison},
but it would be sufficient to mention here that variation of $g^{(2)}(\tau)$
with time delay $\tau\in\left[-T,T\right]$ between detection in two
detectors must have a correlation dip at $\tau=0$ for an ideal single
photon source. 

\subsection{Role in quantum state engineering }

BS plays an important role in quantum state engineering. The field
of generation of desired quantum state by performing different unitary
and non-unitary operations is called quantum state engineering \cite{makhlin2001quantum,malpani2019lower,malpani2019quantum,malpani2019manipulating,malpani2019impact,leonski2011quantum}.
The desired states are usually not available naturally and are required
in many quantum information processing tasks. For example, the output
of the BS, with an input $\left|\psi\right\rangle $ and vacuum $\left|0\right\rangle $
states sent through two inputs ports, can be written as \cite{agarwal2013quantum}
\begin{equation}
\begin{array}{lcl}
\left|\psi_{{\rm out}}\right\rangle  & = & U_{{\rm BS}}\left|\psi\right\rangle \left|0\right\rangle ,\\
 & = & \left|\psi\right\rangle \left|0\right\rangle +i\theta\left(a\left|\psi\right\rangle \right)\left|1\right\rangle ,
\end{array}\label{eq:sub}
\end{equation}
where we have assumed a highly transmitive BS. Notice that conditioned
on a single photon detection in the second output port, a single photon
is subtracted from the input state $\left|\psi\right\rangle $. This
method of photon subtraction is a probabilistic process. Some of these
engineered quantum states are found useful in quantum communication
schemes (\cite{srikara2019continuous,ma2018continuous} and references
therein). Additionally, a BS is an integral optical element for implementation
of quantum scissors in generating finite dimensional nonclassical
states \cite{leonski2011quantum} and entanglement generation \cite{zukowski1997realizable}.

\subsection{Idea of an optical qubit \label{subsec:qubit}}

Assuming one of the inputs of the BS in vacuum and other as single
photon $\left(\left|1\right\rangle \right)$ initially, the output
modes can be described as 
\begin{equation}
\begin{array}{lcl}
\left|\psi_{{\rm 2}}\right\rangle  & = & U_{{\rm BS}}\left|1,0\right\rangle ,\\
 & = & \left(\cos\theta\left|1,0\right\rangle +i\sin\theta\left|0,1\right\rangle \right).
\end{array}\label{eq:qubit}
\end{equation}
In the Fock basis, $\left|\psi_{{\rm 2}}\right\rangle $ can be described
as an Entangled state (for $\theta\neq\frac{n\pi}{2}$ with integer
$n$). Fock (number) basis is the set of orthonormal functions which
are eigen functions of Harmonic oscillator. One can define logical 
bit values $\left|1,0\right\rangle $
as $\left|0\right\rangle _{L}$ and $\left|0,1\right\rangle $ as
$\left|1\right\rangle _{L}$, and thus $\left|\psi_{{\rm 2}}\right\rangle =\left(\cos\theta\left|0\right\rangle _{L}+i\sin\theta\left|1\right\rangle _{L}\right)$
represents an optical qubit in path degree of freedom. This is also
known as a dual-rail qubit. It is noteworthy here that a qubit can be
defined in other degrees of photon as well, such as polarization,
orbital angular momentum. 

This also plays a significant role in introducing the idea of measurement
postulate of quantum mechanics. For instance, $\left|\psi_{{\rm 2}}^{\prime}\right\rangle =\left|\psi_{{\rm 2}}\right\rangle _{\theta=\frac{\pi}{4}}=\frac{1}{\sqrt{2}}\left(\left|0\right\rangle _{L}+i\left|1\right\rangle _{L}\right)$
for a symmetric BS, and a detector on each output port of the BS destroys
this superposition of paths and gives us a single photon detection
at one of the detectors with probability $1/2$ each. This forms the
theoretical basis of commercially available quantum random number
generators \cite{herrero2017quantum} as the randomness is intrinsic
in this case. 

Further, it is straightforward to understand the idea of quantum computation
and MZI/MI in which case a mirror, i.e., $t=0$ and $r=i$ in Eq.
(\ref{eq:BS}), is applied on both $\left|0\right\rangle _{L}$ and
$\left|1\right\rangle _{L}$ in $\left|\psi_{{\rm 2}}\right\rangle $
to result in $i\left|\psi_{{\rm 2}}\right\rangle $. This further
evolves to $\left|\psi_{3}\right\rangle =iU_{{\rm BS}}\left|\psi_{{\rm 2}}\right\rangle =i\left(\cos2\theta\left|0\right\rangle _{L}+i\sin2\theta\left|1\right\rangle _{L}\right)$,
which reduces to $\left|\psi_{3}\right\rangle =-\left|1\right\rangle _{L}$
for the symmetric BS. We have already mentioned that gravitational
wave detection uses an MI with squeezed vacuum inserted through the
second input port of the BS \cite{aasi2013enhanced,grote2013first}.
As an MI is primarily built using a BS and two mirrors, in
view of the above, we can comment that gravitational wave detection
setup in LIGO was essentially built using 3 BSs only.

\subsection{Linear optical Bell state discrimination}

Bell basis has four maximally entangled orthogonal two-qubit states.
Bell states in the polarization degree of freedom can be defined as
$\left|\psi^{\pm}\right\rangle =\frac{1}{\sqrt{2}}\left(a_{H}^{\dagger}b_{H}^{\dagger}\pm a_{V}^{\dagger}b_{V}^{\dagger}\right)\left|0,0\right\rangle $
and $\left|\phi^{\pm}\right\rangle =\frac{1}{\sqrt{2}}\left(a_{H}^{\dagger}b_{V}^{\dagger}\pm a_{V}^{\dagger}b_{H}^{\dagger}\right)\left|0,0\right\rangle ,$
where $\left|0,0\right\rangle $ is the two-mode vacuum state, and
the subscripts represent horizontal $\left(H\right)$ and vertical
$\left(V\right)$ polarization. To understand the idea of Bell measurement
(BM) with linear optics and the role of BS in that, we can apply the
symmetric BS operation on the input Bell state. For instance, on application
of a BS $\left|\psi^{\pm}\right\rangle $ would become 
\begin{equation}
\begin{array}{lcl}
U_{{\rm BS}}\left|\psi^{\pm}\right\rangle  & = & \frac{i}{\sqrt{2}}\left(a_{H}^{\dagger2}+b_{H}^{\dagger2}\pm a_{V}^{\dagger2}\pm b_{V}^{\dagger2}\right)\left|0,0\right\rangle ;\end{array}\label{eq:psi}
\end{equation}
while $\left|\phi^{+}\right\rangle $ and $\left|\phi^{-}\right\rangle $
would be transformed to 

\begin{equation}
\begin{array}{lcl}
U_{{\rm BS}}\left|\phi^{+}\right\rangle  & = & \frac{i}{\sqrt{2}}\left(a_{H}^{\dagger}a_{V}^{\dagger}+b_{H}^{\dagger}b_{V}^{\dagger}\right)\left|0,0\right\rangle \end{array}\label{eq:phip}
\end{equation}
and
\begin{equation}
\begin{array}{lcl}
U_{{\rm BS}}\left|\phi^{-}\right\rangle  & = & \frac{1}{\sqrt{2}}\left(a_{H}^{\dagger}b_{V}^{\dagger}-a_{V}^{\dagger}b_{H}^{\dagger}\right)\left|0,0\right\rangle \end{array},\label{eq:phim}
\end{equation}
respectively. Notice that $\left|\psi^{\pm}\right\rangle $ and $\left|\phi^{+}\right\rangle $
result in both photons in the same output of the BS (cf. Eqs. (\ref{eq:psi})-(\ref{eq:phip})),
whereas only $\left|\phi^{-}\right\rangle $ gives one photon in each
output port of the BS (in Eq. (\ref{eq:phim})). This behavior can
be attributed to the fact that singlet state shows fermionic behavior
at a BS, while rest of the triplet Bell states show bosonic
nature \cite{weihs2001photon}. Therefore, a BS is able to identify
one of the Bell states, i.e., $\left|\phi^{-}\right\rangle $, out
of total four Bell states. 

From Eqs. (\ref{eq:psi})-(\ref{eq:phip}) it can be observed that
both photons in one of the output ports of the BS have same (orthogonal)
polarization for Bell state $\left|\psi^{\pm}\right\rangle $ $\left(\left|\phi^{+}\right\rangle \right)$.
Exploiting this fact, we can further identify $\left|\phi^{+}\right\rangle $
if we place a polarizing beamsplitter (PBS) at both the outputs of
the BS (as shown in Fig. \ref{fig:BS} (c)). Unlike a polarization
independent BS introduced in Eq. (\ref{eq:BS}), a PBS is a particular
type of polarization dependent BS which reflects (transmits) vertically
(horizontally) polarized photons. Thus, a photon is detected each
at ${\rm D}_{{\rm 1V}}$ and ${\rm D}_{{\rm 2H}}$ (or ${\rm D}_{{\rm 3V}}$
and ${\rm D}_{{\rm 4H}}$) in Fig. \ref{fig:BS} (c) for $\left|\phi^{-}\right\rangle $,
whereas a photon is detected each at ${\rm D}_{{\rm 1V}}$ and ${\rm D}_{{\rm 4H}}$
(or ${\rm D}_{{\rm 3V}}$ and ${\rm D}_{{\rm 2H}}$) for $\left|\phi^{+}\right\rangle $.

It is noteworthy that a single BS is sufficient to identify one of
the Bell states, while a single PBS can be used to check parity of
the Bell states which is useful in quantum error correction codes
\cite{pathak2013elements}. 

\section{Beamsplitter in discrete variable quantum communication \label{sec:discrete}}

Using the optical resources and photon number measurements discussed
in Section \ref{sec:Mathematical-modeling}, we will briefly introduce
discrete variable insecure and secure quantum communication.

\subsection{Quantum teleportation \label{subsec:Quantum-teleportationdv}}

\begin{figure}
\begin{centering}
\includegraphics{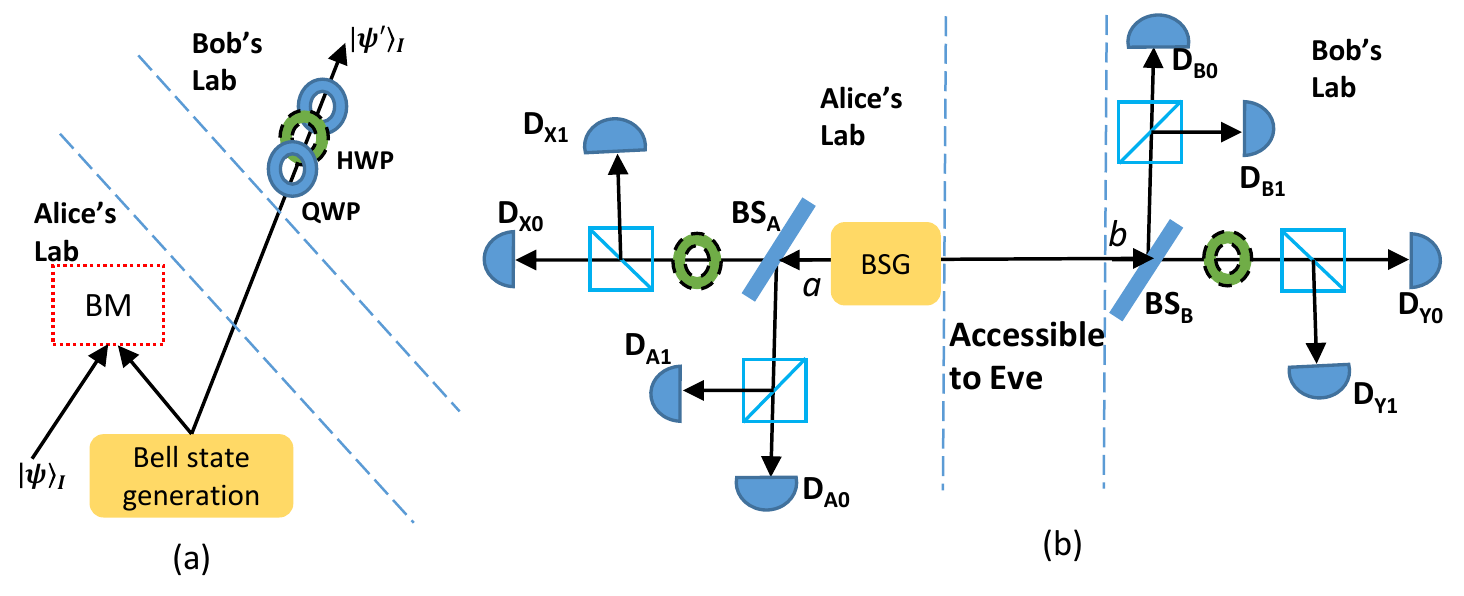}
\par\end{centering}
\caption{\label{fig:DV}(Color online) (a) Teleportation of $\left|\psi\right\rangle _{I}$
using BM shown in Fig. \ref{fig:BS}, quarter (QWP) and half (HWP)
wave-plates as well as Bell state generation (BSG). (b) Quantum key
distribution using BSG, BSs, PBSs, and HWPs.}
\end{figure}

We may now discuss the teleportation \cite{bennett1993teleporting}
of a qubit $\left|\psi\right\rangle _{I}=\left(\alpha c_{H}^{\dagger}+\beta c_{V}^{\dagger}\right)\left|0\right\rangle $
with the help of shared bipartite quantum channel $\left|\phi^{-}\right\rangle $
between sender Alice and receiver Bob. The combined state of channel,
after passing the qubit to be transmitted and Alice's part of the
bipartite state through two input ports of the BS, is 
\begin{equation}
\begin{array}{lcl}
\left|\psi^{\prime}\right\rangle  & = & \left(U_{{\rm BS}}\otimes I_{2}\right)\left|\psi\right\rangle _{I}\otimes\left|\phi^{-}\right\rangle \\
 & = & \frac{1}{2\sqrt{2}}\left\{ i\alpha\left(c_{H}^{\dagger2}+a_{H}^{\dagger2}\right)b_{V}^{\dagger}+\beta\left(ic_{H}^{\dagger}c_{V}^{\dagger}+ia_{H}^{\dagger}a_{V}^{\dagger}-c_{H}^{\dagger}a_{V}^{\dagger}+c_{V}^{\dagger}a_{H}^{\dagger}\right)b_{V}^{\dagger}\right.\\
 & - & \left.\alpha\left(ic_{H}^{\dagger}c_{V}^{\dagger}+ia_{H}^{\dagger}a_{V}^{\dagger}+c_{H}^{\dagger}a_{V}^{\dagger}-c_{V}^{\dagger}a_{H}^{\dagger}\right)b_{H}^{\dagger}-i\beta\left(c_{V}^{\dagger2}+a_{V}^{\dagger2}\right)b_{H}^{\dagger}\right\} \left|0,0,0\right\rangle .
\end{array}\label{eq:tel}
\end{equation}
Here, $I_{2}$ is a $2\times2$ identity matrix. Using the linear
optical Bell state measurement discussed previously, we discard the
first and last cases in Eq. (\ref{eq:tel}) and write the unnormalized
state as 
\begin{equation}
\begin{array}{lcl}
\left|\psi^{\prime}\right\rangle  & = & -\frac{1}{2\sqrt{2}}\left\{ \left(c_{H}^{\dagger}a_{V}^{\dagger}-c_{V}^{\dagger}a_{H}^{\dagger}\right)\left(\alpha b_{H}^{\dagger}+\beta b_{V}^{\dagger}\right)+i\left(c_{H}^{\dagger}c_{V}^{\dagger}+a_{H}^{\dagger}a_{V}^{\dagger}\right)\left(\alpha b_{H}^{\dagger}-\beta b_{V}^{\dagger}\right)+D\right\} \left|0,0,0\right\rangle .\end{array}\label{eq:tel2}
\end{equation}
where $D$ represents the rest of the cases which are discarded. Here,
the first (second) term corresponds to Alice's BM result $\left|\phi^{-}\right\rangle $
$\left(\left|\phi^{+}\right\rangle \right)$ and thus requires identity
(Pauli $Z$) operation on Bob's qubit to reconstruct teleported state
as $\left|\psi^{\prime}\right\rangle _{I}$. Here, Pauli $Z$ operation
can be performed by using waveplates. Schematic diagram of the quantum
teleportation is shown in Fig. \ref{fig:DV} (a). A number of variants
of quantum teleportation scheme \cite{shukla2016hierarchical,sisodia2017design,thapliyal2015applications,thapliyal2015general}
are theoretically proposed in the past, which will require BS for
experimental implementation.

\subsection{Quantum cryptography\label{subsec:Quantum-cryptographydv}}

The idea of quantum cryptography can be understood by a quantum key
distribution scheme \cite{bennett1984quantum,bennett1992B92}. To
begin with Alice (sender) prepares a string of an entangled state
$\left|\psi^{+}\right\rangle $ and shares the second mode (qubit
in this case) with Bob. Both Alice and Bob pass their individual modes
through a BS, namely ${\rm BS}_{{\rm A}}$ and ${\rm BS}_{{\rm B}}$, respectively.
Therefore, the combined state $\left|\psi^{+}\right\rangle $ is transformed
to 
\begin{equation}
\begin{array}{lcl}
\left|\psi^{\prime}\right\rangle  & = & \frac{1}{2\sqrt{2}}\left(A_{H}^{\dagger}B_{H}^{\dagger}+A_{V}^{\dagger}B_{V}^{\dagger}-X_{H}^{\dagger}Y_{H}^{\dagger}-X_{V}^{\dagger}Y_{V}^{\dagger}+i\left\{ A_{H}^{\dagger}Y_{H}^{\dagger}+A_{V}^{\dagger}Y_{V}^{\dagger}+X_{H}^{\dagger}B_{H}^{\dagger}+X_{V}^{\dagger}B_{V}^{\dagger}\right\} \right)\left|0,0\right\rangle ,\end{array}\label{eq:cry}
\end{equation}
which can be further transformed by applying a quarter waveplate on
mode $X$ and $Y$ to transform photons in the rectilinear basis to
diagonal basis (i.e., $H\rightarrow\frac{H+V}{\sqrt{2}}$ and $V\rightarrow\frac{H-V}{\sqrt{2}}$)
as 
\begin{equation}
\begin{array}{lcl}
\left|\psi^{\prime\prime}\right\rangle  & = & \frac{1}{2\sqrt{2}}\left(A_{0}^{\dagger}B_{0}^{\dagger}+A_{1}^{\dagger}B_{1}^{\dagger}-X_{0}^{\dagger}Y_{0}^{\dagger}-X_{1}^{\dagger}Y_{1}^{\dagger}+D\right)\left|0,0\right\rangle .\end{array}\label{eq:cry-2}
\end{equation}
Here, we have written $H\equiv0,$ $V\equiv1,$ $\frac{H+V}{\sqrt{2}}\equiv0,$
and $\frac{H-V}{\sqrt{2}}\equiv1$; and rest of the cases are discarded
(shown as $D$). Notice that in all the cases which are not discarded,
Alice's and Bob's bit values are symmetric. Therefore, this scheme
enables Alice and Bob to share a symmetric key, which provides security
not conditioned on some computationally complex problem like classical
cryptography. Schematic diagram of this quantum key distribution scheme
is shown in Fig. \ref{fig:DV} (b). Both Alice and
Bob check half of the obtained string to ensure that an adversary
Eve has not tried to eavesdrop, which would have left detectable traces
in the form of errors in the measurement outcomes. The security is
further enhanced by error correction and privacy amplification. 

The present scheme is same as quantum cryptography scheme proposed
in \cite{bennett1992quantum}. It is possible for Alice to measure
her qubit in $\left|\psi^{+}\right\rangle $ before sending the second
qubit to Bob, the present scheme reduces to the first quantum cryptography
scheme, BB84 scheme, proposed by Bennett and Brassard \cite{bennett1984quantum}. 

This idea can be further extended to measurement device independent
quantum key distribution scheme \cite{lo2012measurement} where both
Alice and Bob prepare a string analogous to BB84 scheme and send to
a third party Charlie, midway between Alice and Bob. Charlie performs
BM as described in Section \ref{sec:discrete} and announces the successful
cases of measurement outcomes $\left|\phi^{+}\right\rangle $ and
$\left|\phi^{-}\right\rangle $. These two cases correspond to orthogonal
states prepared by Alice and Bob, and thus Alice and Bob obtain a
symmetric key once Bob flips all the bit values in his key. 

\section{Beamsplitter in continuous variable quantum communication \label{sec:continuous}}

Homodyne/Heterodyne measurement, instead of single photon detectors
in the discrete variable communication schemes (in Section \ref{sec:discrete}),
is central idea for continuous variable communication. There are certain
advantages of this type of quantum communication as it allows one
to use existing optical technology to perform metropolitan quantum
communication, which exempts us from expensive single photon source
and detector (\cite{srikara2019continuous,saxena2019continuous} and
references therein). 

\subsection{Quantum teleportation\label{subsec:Quantum-teleportationcv}}

The idea of continuous variable quantum teleportation \cite{braunstein1998teleportation}
is analogous to that described in Section \ref{subsec:Quantum-teleportationdv}.
As probability amplitudes of a quantum state are transferred in discrete
variable teleportation, canonically conjugate continuous variable
quadratures $x_{{\rm in}}$ and $p_{{\rm in}}$ of an unknown coherent
state are transmitted here. Alice and Bob are expected to share bipartite
continuous variable entanglement with quadrature variables $\left(x_{A},p_{A}\right)$
and $\left(x_{B},p_{B}\right)$, respectively. Alice passes the mode
of bipartite entanglement and unknown coherent state through a symmetric
BS and measures quadratures $x_{{\rm in}}^{\prime}=\frac{1}{\sqrt{2}}\left(x_{{\rm in}}+x_{A}\right)$
and $p_{A}^{\prime}=\frac{1}{\sqrt{2}}\left(p_{{\rm in}}-p_{A}\right)$
in each output of the BS to perform BM. Subsequently, she announces
the measurement outcomes $\bar{x}$ and $\bar{p}$ to Bob, who performs
displacement operator to obtain $x_{{\rm out}}=\left(x_{B}+\mathcal{G}\bar{x}\right)$
and $p_{{\rm out}}=\left(p_{B}+\mathcal{G}\bar{p}\right)$ with gain
factor $\mathcal{G}=\sqrt{2}$. Notice that $x_{{\rm out}}=x_{{\rm in}}+\left(x_{A}+x_{B}\right)$
and $p_{{\rm out}}=p_{{\rm in}}-\left(p_{A}-p_{B}\right)$ therefore
for a perfect teleportation of continuous quantum variables the initial
bipartite entanglement Alice and Bob share should minimize the noise
$\left(x_{A}+x_{B}\right)$ and $\left(p_{A}-p_{B}\right)$. This
property can be satisfied by two-mode squeezed vacuum state as $\left\langle \left(x_{A}+x_{B}\right)^{2}\right\rangle =\left\langle \left(p_{A}-p_{B}\right)^{2}\right\rangle =\exp\left(-2r\right)$
which tends to zero in case of infinitely strong squeezing, i.e.,
$r\rightarrow\infty$.

Interestingly, a complete description of $n$-mode Gaussian states
(a state fully characterized by its first and second moments only)
can be provided by corresponding $2n$ dimensional covariance matrix
$\sigma$ \cite{weedbrook2012gaussian} with $\sigma_{ij}=\left\langle \left\{ \Delta R_{i},\Delta R_{j}\right\} _{+}\right\rangle $,
where $\left\{ A,B\right\} _{+}=\frac{1}{2}\left(AB+BA\right)$. Here,
the vector $R=\left(x_{1},p_{1},\ldots x_{n},p_{n}\right)^{T}$ is
defined in terms of quadrature variables ensuring $\left[R_{j},R_{k}\right]=i\Omega_{jk}$
with $\Omega=\stackrel[k=1]{n}{\oplus}\omega$ and $\omega=\left(\begin{array}{cc}
0 & 1\\
-1 & 0
\end{array}\right),$ and $\Delta R_{i}=R_{i}-\left\langle R_{i}\right\rangle $. Necessary
and sufficient condition for a matrix to be a covariance matrix based
on uncertainty relation is $\sigma+i\Omega>0$.

Thus, teleportation of a Gaussian state with covariance matrix $\sigma_{{\rm in}}$
can be performed by using prior shared entanglement $\sigma_{AB}$
\cite{mista2005improving}. Alice performs BM by using a BS. The BS
(in general, any unitary) operation is represented by a symplectic
transformation $S_{{\rm BS}}=\left(\begin{array}{cc}
\cos\theta I_{2} & \sin\theta S_{P}\\
-\sin\theta S_{P}^{T} & \cos\theta I_{2}
\end{array}\right)$ with symplectic matrix for phase shift $S_{P}=\left(\begin{array}{cc}
\cos\Phi & \sin\Phi\\
-\sin\Phi & \cos\Phi
\end{array}\right)$, which satisfies $S_{{\rm BS}}\Omega S_{{\rm BS}}^{T}=\Omega$ analogous
to unitary condition. Following the teleportation scheme (discussed
above) with $\theta=\frac{\pi}{4}$ and $\Phi=\frac{\pi}{2}$, Bob
obtains $R_{{\rm out}}=R_{B}+\mathcal{G}\bar{R}$ and corresponding
covariance matrix $\sigma_{{\rm out}}=\sigma_{{\rm in}}+2N$, where
the additional term $2N$ is noise introduced \cite{mista2005improving}.
Assuming the Gaussian state to be teleported as a coherent state $\sigma_{{\rm in}}=\frac{1}{2}I_{2}$,
and two-mode squeezed vacuum state as shared channel $\sigma_{AB}=\left(\begin{array}{cc}
A & C\\
C^{T} & B
\end{array}\right)$ with $A=B=\frac{1}{2}\cosh\left(2r\right)I_{2}$ and $C={\rm diag}\left(-\frac{1}{2}\sinh\left(2r\right),\frac{1}{2}\sinh\left(2r\right)\right)$.
In that case, the noise is minimum, i.e., $2N=\exp\left(-2r\right)I_{2}$
which becomes zero for $r\rightarrow\infty$. Two-mode squeezed vacuum
state is an example of Einstein-Podolsky-Rosen (EPR) entanglement
\cite{braunstein1998teleportation}, which can be generated by sending
two single-mode states equally squeezed in different (say $x_{a}$
and $p_{b}$) quadratures through two input ports of the BS. Advantage
in the performance of continuous variable teleportation is proposed
by using local squeezing operations on the bipartite entanglement
shared by Alice and Bob \cite{mista2005improving}.

\subsection{Quantum cryptography\label{subsec:Quantum-cryptographycv}}

Analogous to the discrete variable quantum cryptography scheme discussed
in Section \ref{subsec:Quantum-cryptographydv}, Alice can prepare
and share two-mode squeezed vacuum state with covariance matrix $\sigma_{AB}$
with Bob. Alice and Bob can measure one of the quadratures on their
part of the state and thus obtain a string of bits corresponding to
measurement outcomes when they measured the same quadrature using
homodyne technique. They check their measurement outcomes in one-half
of these cases to check and if they are not correlated it can be attributed
to the eavesdropping attempts by Eve. Thus, using error correction
and privacy amplification a secure key can be generated. However,
note that a continuous quantum variable is used to encode a discrete
quantum key in this case and thus this type of quantum key distribution
schemes is
categorized
as hybrid continuous variable quantum key distribution schemes \cite{cerf2001quantum}. 

More recently quantum key distribution schemes are presented where
Alice performs a single mode squeezing operation $s$ on the mode
sent to Bob \cite{cerf2001quantum,jacobsen2018complete}. Thus, $\sigma_{AB}$
transforms to $\sigma_{AB}^{\prime}$ with $B={\rm diag}\left(\frac{1}{2}\cosh\left(2r\right)e^{-2s},\frac{1}{2}\cosh\left(2r\right)e^{2s}\right)$
and $C={\rm diag}\left(-\frac{1}{2}\sinh\left(2r\right)\exp\left(-s\right),\frac{1}{2}\sinh\left(2r\right)\exp\left(s\right)\right)$.
After Alice's measurement of quadrature $x_{A}$ or $p_{A}$ the reduced
covariance matrix for Bob can be obtained as $\sigma_{B_{x}}^{\prime\prime}=B-2{\rm sech}\left(2r\right)C^{T}\Pi C$
or $\sigma_{B_{p}}^{\prime\prime}=B-2{\rm sech}\left(2r\right)C^{T}\Pi^{\prime}C$,
respectively, with $\Pi=\left(\begin{array}{cc}
1 & 0\\
0 & 0
\end{array}\right)$ and $\Pi^{\prime}=\left(\begin{array}{cc}
0 & 0\\
0 & 1
\end{array}\right).$

BS and homodyne detection are also required in continuous variable
quantum key distribution using non-Gaussian channels \cite{srikara2019continuous}
and direct secure quantum communication (which allow us to perform
secure quantum communication without generating and distributing a
quantum key) schemes \cite{saxena2019continuous}.

\begin{figure}
\begin{centering}
\includegraphics{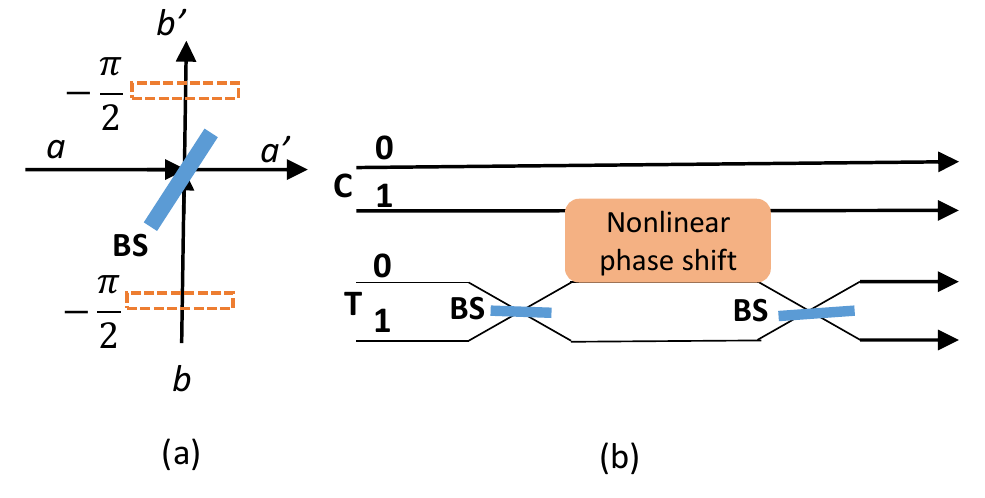}
\par\end{centering}
\caption{\label{fig:comp}(Color online) (a) Hadamard using a BS and phase
plates. (b) CNOT using two BSs and nonlinear phase shift with C and
T corresponding to control and target qubits, respectively.}
\end{figure}

\section{Beamsplitter in discrete and continuous variable optical quantum
computation\label{sec:computation}}

Quantum computation using linear optical resources can be performed
using BS, phase shifter, and mirror with single photon detectors and
quantum memory \cite{kok2007linear,o2007optical}. While introducing
the idea of qubit in Section \ref{subsec:qubit}, we have already shown
that $U_{{\rm BS}}\left|0\right\rangle _{L}=\left(\cos\theta\left|0\right\rangle _{L}+i\sin\theta\left|1\right\rangle _{L}\right)$,
which reduces to $U_{{\rm BS}}\left|0\right\rangle _{L}=\frac{1}{\sqrt{2}}\left(\left|0\right\rangle _{L}+i\left|1\right\rangle _{L}\right)$
for symmetric BS. Thus, a BS and two phase shifters are sufficient
to perform hadamard gate \cite{kok2007linear} (shown in Fig. \ref{fig:comp}
(a)). However, quantum computing requires feasibility of library of
universal quantum gates, preparation of the initial quantum states,
and measurement of the final state. Hong-Ou-Mandel effect \cite{hong1987measurement},
i.e., two indistinguishable single photons mixing at a symmetric BS
$U_{{\rm BS}}a^{\dagger}b^{\dagger}\left|0,0\right\rangle =\frac{i}{\sqrt{2}}\left(a^{\dagger2}+b^{\dagger2}\right)\left|0,0\right\rangle $
coalesce to the same output arm of the BS, plays an important role
in the implementation of two-qubit gates in quantum computing. For
instance, CNOT gate on the spatial qubits uses two BSs, to employ
Hong-Ou-Mandel effect, and a controlled phase gate \cite{o2007optical}
as shown in Fig. \ref{fig:comp} (b). Therefore, a set of universal
quantum gates \cite{nielsen2010quantum}, including Hadamard, phase
and CNOT gates, can be performed using BS. Further, an $n$-port unitary
can be implemented by phase shifters and only $n\left(n-1\right)/2$
BSs \cite{kok2007linear}. Similarly, BS is relevant in designing
several other optical gates, like controlled phase and nonlinear sign
gates as well as CNOT with polarization qubits and hyperentanglement
(see \cite{kok2007linear,ralph2010optical} for detail). CNOT with
optical fiber is also implemented experimentally \cite{o2003demonstration}.
Significant contributions in the field were performed using ancilla
photons by KLM approach \cite{knill2000efficient}, which was further
improved in \cite{franson2002high}.

Similarly, BS plays an important role in continuous variable quantum
computation \cite{ralph2003quantum,gu2009quantum}, we refrain us
from discussing it further here.

\section{Other applications in the field of quantum optics and technology
\label{sec:Other}}

Significance of BS in several aspects of quantum optical and information
processing experiments is difficult to summarize in this article.
Therefore, here we briefly mention some of these applications of BS
in entanglement concentration protocols \cite{tatham2014entanglement},
quantum repeaters \cite{simon2007quantum}, quantum simulation \cite{aspuru2012photonic},
cryptanalysis in quantum cryptography \cite{pirandola2008continuous,liu2011proof},
linear optical coupler--equivalent to BS operation--is found relevant
in the study of non-Hermitian physics or parity-time symmetry \cite{naikoo2019quantum},
implementation of quantum cryptography \cite{sibson2017chip}, computation
\cite{o2009photonic}, and technology \cite{o2009photonic}, etc.
Further, BS, as an ingredient of MZI and MI, is used in the studies
of quantum Zeno effect \cite{agarwal1994all} and its use in counterfactual
quantum communication \cite{noh2009counterfactual,brida2012experimental,cao2017direct}
and computation \cite{hosten2006counterfactual,kong2015experimental};
Elitzur-Vaidman bomb testing or interaction free measurement \cite{elitzur1993quantum}
which is useful in Guo-Shi \cite{guo1999quantum} quantum cryptography
scheme; Goldenberg-Vaidman quantum key distribution \cite{goldenberg1995quantum},
quantum phase estimation \cite{higgins2007entanglement,kacprowicz2010experimental}
used in quantum metrology \cite{giovannetti2011advances} and quantum
radar \cite{lanzagorta2011quantum}; experiments relevant in foundations
of quantum mechanics \cite{dhand2018understanding} as delayed choice
measurement \cite{polino2019device}, realization of Hardy's paradox
\cite{irvine2005realization}, wave-particle duality \cite{grangier1986experimental,li2017weak},
violation of Bell's inequality \cite{rarity1990experimental}, device
independence \cite{polino2019device,gisin2010proposal}, and weak
measurements \cite{li2017weak}; and gravitational wave detection
\cite{aasi2013enhanced,grote2013first}.

As far as the Hamiltonian (\ref{eq:ham}) is concerned, it describes
Bose-Einstein condensates \cite{mewes1997output,torii2000mach}, optomechanical
systems \cite{aspelmeyer2014cavity}, and plasmonic circuits \cite{heeres2013quantum}
as well. Additionally, evolution after taking into consideration weak
nonlinearity for BS, optical coupler or other physical systems is
also studied in the past \cite{giri2017nonclassicality,thapliyal2014higher,thapliyal2014nonclassical,thapliyal2016linear,mukhopadhyay2019interaction}.
The applications are further extended to slow light beam splitters
as well \cite{xiao2008slow}. 

\section{Summary and concluding remarks\label{sec:Conclusion}}

The dynamical evolution of the quantum state of a quantum system plays
a significant role in quantum mechanics and its experiments. Thus,
all the optical elements used in experimental implementation are represented
by unitary operations. One of the most important Hamiltonians, often
used in quantum optics and information processing, governing dynamics
of the optical states (defined using different properties of photon,
such as polarization, frequency, orbital angular momentum) is BS operation.
Here, we discuss in detail the significance of BS operation in experimental
studies ranging from foundational verification of principles of quantum
mechanics to quantum optics, quantum information processing, and technology. 

Specifically, BS is useful in characterization of nonclassical--antibunched,
squeezed, entangled, steerable, Bell nonlocal--states, studies of
higher-order nonclassicality, measurement of continuous variable quantum
states, quantum state engineering for photon subtraction and entanglement
generation, linear optical Bell state discrimination, discrete and
continuous variable quantum teleportation and cryptography, cryptanalysis
of secure quantum communication schemes, discrete and continuous variable
quantum computation, quantum phase estimation, gravitational wave
detection to name a few. However, there are some limitations of BS
operation, for instance, generation of entanglement using classical
resources, optical CNOT, deterministic Bell state discrimination cannot
be performed using linear optics solely. 

In brief, in the present work, we have tried to reveal the inherent
symmetry present in many physical processes of relevance and interest.
The inherent symmetry is investigated here by using one of the simplest
possible optical components (BS). This investigation is performed from
a new approach, and it is expected to complement a set of earlier
studies \cite{henault2015quantum,holbrow2002photon,zeilinger1981general,weihs2001photon}
focused on properties and applications of BS. This article is also
expected to be of use in teaching/training young students about the
relation between optics, quantum mechanics, and quantum information.
If it succeeds in that then that would be our greatest possible tribute
to Prof.\ Ghatak who has spent most part of his life in writing books
and articles for young students with a clear focus on clarifying complex
ideas in a lucid manner.

It is fascinating to observe that an optical element which was known
and used in some form or others in the early civilizations, is still
used to produce new results and to obtain new insights into the physical
world. We hope the journey will continue and BS-type simple physical
systems will continue to help us in enriching our understanding of
the nature. Keeping the earlier stated points and this hope in mind,
we conclude this article by noting (in analogy with Keats) that simple
(BS) is beauty, and beauty is truth.


\end{document}